\documentclass{emulateapj}
\usepackage{amsmath}
\usepackage{natbib}
\usepackage[caption=false,position=top,labelformat=empty]{subfig}
\begin{document}
\title{Preferential heating and acceleration of Heavy Ions in Impulsive Solar Flares}

\author{Rahul Kumar$^{1,2}$, David Eichler$^{2}$,  Massimo Gaspari$^{1,3}$, Anatoly Spitkovsky$^{1}$}
\altaffiltext{1}{Department of Astrophysical Sciences, Princeton University, Princeton, NJ 08544, USA}
\altaffiltext{2}{Department of Physics, Ben-Gurion University, Be'er-Sheba 84105, Israel}
\altaffiltext{3}{Einstein and Spitzer Fellow}
\date{\today}
\begin{abstract}
We simulate decaying turbulence in a homogeneous pair plasma using three dimensional electromagnetic particle-in-cell(PIC) method. A uniform background magnetic field permeates the plasma such that the magnetic pressure is three times larger than the thermal pressure and the turbulence is generated by counter-propagating shear Alfven waves. The energy predominately cascades transverse to the background magnetic field, rendering the turbulence anisotropic at smaller scales. We simultaneously move several ion species of varying charge to mass ratios in our simulation and show that the particles of smaller charge to mass ratios are heated and accelerated to non-thermal energies at a faster rate, in accordance with the enhancement of heavy ions and non-thermal tail in their energy spectrum observed in the impulsive solar flares. We further show that the heavy ions are energized mostly in the direction perpendicular to the background magnetic field with a rate consistent with our analytical estimate of the rate of heating due to cyclotron resonance with the Alfven waves of which a large fraction is due to obliquely propagating waves. 
\end{abstract}

\section{Introduction}

The ceaseless stream of ions and electrons originating from the sun often shows short-term enhancement in the flux of energetic charged particles, known as the solar energetic particles (SEPs). While long duration gradual SEP events lasting about few days are associated with shock waves due to coronal mass ejections(CMEs), impulsive SEP events lasting several hours appear to have originated from compact regions near the sun which are associated with the flare events \citep{Reames2015}. Impulsive SEPs are often characterized by greatly enhanced $^3He/^4He$ ratio over the solar abundance ratio. Measurements of the flux of particles in differential energy bins suggest that a fraction of particles are accelerated to very high energies which form a non-thermal tail in energy distribution \citep{Mason2002,Mason2007}. Additionally, $^3He$ rich SEPs show enhancement of other heavy ions as well, and the enhancement factor is generally larger for the heavier ions \citep{Mason2004}. 

The physical mechanisms that give rise to preferential acceleration of heavier ions in impulsive SEPs are not well understood, although a wide range of observations have set some constraints on the environment where the acceleration takes place. The strong dependence of enhancement on charge to mass ratio, Q/M, of ions indicates that a Q/M dependent acceleration process is at play which disfavors processes such as the diffusive acceleration in shocks where the rate of acceleration depends on rigidity of charged particles (see, however \cite{ShockHeating}). It has been suggested that the cyclotron damping of Alfvenic turbulence by the tail of particle distribution may lead to preferential acceleration of particles with smaller Q/M and larger velocity \citep{Eichler1979,David2014}. Particles of smaller Q/M have smaller gyro-frequency and would therefore resonate with waves of smaller wavenumber. If the power spectrum is a sufficiently steep function of parallel wavenumber, it would imply a larger heating and acceleration rate for smaller Q/M particles. Moreover, faster moving super-Alfvenic particles can resonate with even smaller frequency waves since in the frame of particles the frequency of Alfven waves is  Doppler shifted due to their parallel velocity along the magnetic field lines. Therefore, for super-Alfvenic particles energy gain in one gyration time is an increasing function of particle velocity which can potentially give rise to a power-law tail in velocity distribution \citep{Eichler1979,David2014}. On the other hand acceleration due to cyclotron resonance is believed to be inefficient due to anisotropic nature of the turbulence cascade which creates turbulent eddies elongated along the large scale magnetic field \citep[see e.g., ][]{Strauss1976,Montgomery1981,Shebalin1983,Chandran2000PRL,Chandran2000,David2014}. Nevertheless, it has been argued that there may be enough power in the Alfven waves propagating along or obliquely to the mean magnetic field to produce observed enhancement of heavier nuclei. 

In this paper we numerically study relative heating and acceleration rate of plasma species of varying Q/M in a Alfvenic turbulence. The ion species heavier than simulation ions (positrons in the case of pair plasma) are treated as test particles in our simulations\citep[see e.g.,][]{Dmitruk2006,Lehe2009,Markovskii2010,Lynn2012,Servidio2015,Gonz2016} and our electromagnetic PIC simulations resolve kinetic scales of all species involved. We show that the rate of energy gain is strongly dependent on Q/M. We then suggest that resonant damping of Alfvenic turbulence is a plausible scenario of preferential heating and acceleration of heavy ions as observed in He3 rich SEPs.

\section{Numerical Simulation} 
We use fully electromagnetic particle-in-cell (PIC) method \citep{Buneman,Rahul} to simulate the development of turbulence in a magnetized collisonless plasma in three spatial dimensions. The simulation box contains uniformly distributed thermal plasma embedded in a uniform Cartesian grid with a background magnetic field $B_0$ along the x-axis. Periodic boundary conditions are imposed along all three cartesian axes. The simulation is started by exciting six shear Alfven waves with wave vector $\vec{k}=(k_x, k_y, k_z)$. Specifically, the initial wave vectors are chosen such that the triplet $(k_x/(2\pi/L_x), k_y/(2\pi/L_y), k_z/(2\pi/L_z))$, where $L_x$, $L_y$, and $L_z$ are size of the box along x, y, and z axes, respectively, takes the values (1,1,0); (1,2,0); (-2,1,0); (-1, 0, -1); (-1, 0, -2); (2, 0, -1). The energy is equally partitioned among these waves and the amplitude of magnetic fluctuation is taken to be $0.18 B_0$  for each wave. The value of initial electromagnetic field (other than the background magnetic field $B_0$) on the grid is superposition of local transverse perturbations corresponding to each initial Alfven modes. The velocity distribution of the charged particles is a drifting Maxwellian where the drift of a particles is the superposition of local $E\times B$ and Polarization drift due to each Alfven modes, where E and B are local value of electric and magnetic field at the particle's initial location. 

In addition to the simulation particles we inject heavier ion species in our simulation (few percent of the simulation particles by number) at the beginning of the simulation to study simultaneous evolution of velocity distribution of ion species characterized by differing Q/M and initial temperature. Heavier ions are treated as test particles. That is to say, heavy ions do not contribute any current in the Maxwell's equations but they are subject to the Lorentz force due to electric and magnetic fields like the simulation particles. The assumption is justified as long as number density of heavy ions is small or damping of turbulence due to heavy ions is small enough that energy budget is not monopolized by them. Like simulation particles, the heavy ions are initially homogeneously distributed in the simulation box and their velocity distribution is a drifting Maxwellian. We initialize several populations of heavy ions with different charge to mass ratio and initial temperatures. 

In order to accommodate large resonant scales of heavy ions in the simulation box while simultaneously resolving the kinetic scales of plasma we choose to simulate electron-positron plasma ( although we still refer to positrons as simulation ions). The physical dimension of the box $L_x$, $L_y$, and $L_z$ in units of electron skin depth are 128, 64, and 64, respectively. Electron skin depth $c/\omega_{pe}$ is resolved by 8 grid cells along each axes, where c is the speed of light and $\omega_{pe}=\sqrt{4\pi n_0 q_e^2/m^r_e}$, where $q_e$ is the charge of an electron, $n_0$ is the initial number density of electrons, and $m^r_e$ is the reduced mass of an electron (for electron-positron plasma considered here reduced mass of each particle is half of its physical mass). On average, there are 12 macro-particles of electrons and ions per unit cell. The Aflven speed $V_A=B_0/\sqrt{4\pi\rho}=0.1c $, where $\rho$ is total initial mass density. The thermal speed of particle is such that the plasma $\beta$, which is ratio of plasma thermal pressure to magnetic pressure, is 1/3. The plasma beta in the environments where impulsive SEPs are produced is likely much larger than 1. A large plasma beta in 3D simulations requires larger simulation box size and computational time to capture cyclotron resonance. However, we substantiate our results with analytical estimates which can be extended to more realistic physical parameters. In other coronal and heliospheric environments the beta parameter may vary over even larger range (at least 0.01 to 10). A numerical study of the dependence of heating and acceleration mechanism on the beta parameter would be attempted in future simulations. 
\section{Turbulence Cascade}

The non-linear interactions among Alfven waves propagating in opposite directions distorts each wave packet, hence generating waves of larger wave numbers \citep[see e.g.,][and the references therein]{Kraichnan1965,Howes2013}. The cascade of energy from larger to smaller length scales rapidly increases power in the waves of larger wave numbers, reaching a near maximum at $k \sim \omega_{pe}/c$ in one Alfven crossing time $\tau_A=L_x/V_A=128/ \Omega_i$, where $\Omega_i$ is the ion (positron) cyclotron frequency ($\Omega_i=q_iB_0/m_i c$, where $q_i$ and $m_i$ are the charge and mass of the simulation ion, respectively). In units of  As the turbulent energy cascades reaches smaller spatial scales electromagnetic field fluctuations are eventually converted into the plasma heat. The amplitude of the large scale Alfven waves decreases with time, since the turbulence is not driven (the total energy in the simulation remains almost constant and does not change by more than 0.1 percent until  $1.5 \tau_A$). Therefore, the turbulence simulated here does not achieve a steady state turbulence cascade. 

\begin{figure}[h]
  \centering
  \includegraphics[width=0.5\textwidth]{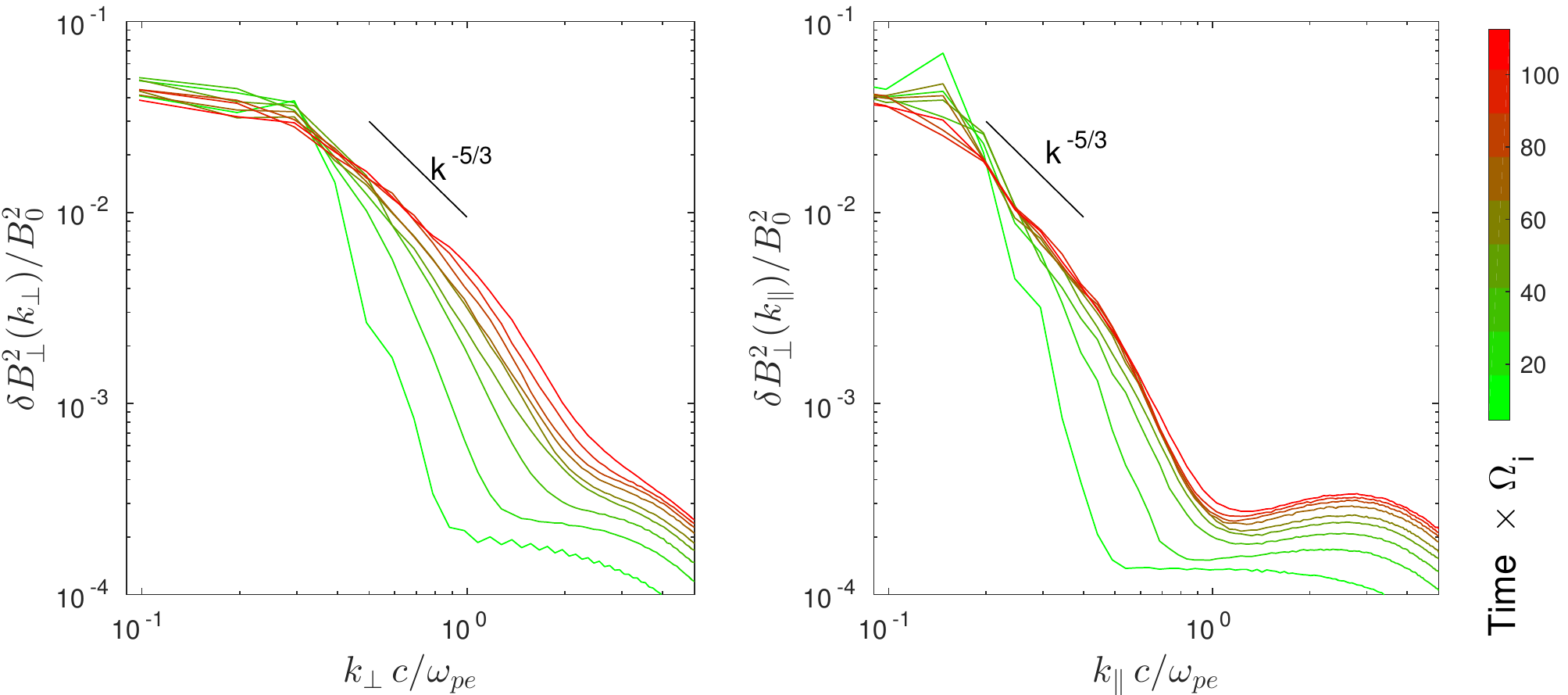}
  \caption{ The curves in the left and right panels show one-dimensional spectrum of transverse magnetic field fluctuation $\delta B_\perp^2=B_y^2+B_z^2$ along the perpendicular and parallel directions, respectively. Each curve shows spectrum at evenly spaced time steps denoted by its color. The spectra show anisotropic nature of the energy cascade.}  
\label{spectrum}
\end{figure}

In figure \ref{spectrum} we show the temporal evolution of one-dimensional spectrum of transverse magnetic field along parallel and perpendicular directions with respect to the background magnetic field. As evident from the spectra, the turbulent cascade of magnetic energy is anisotropic \citep{Goldreich1995,Schekochihin2009,Wan2015}. The energy cascades preferentially in the direction perpendicular to the background magnetic field and creates elongated eddies. 

\begin{figure}[h]
  \includegraphics[width=0.45\textwidth]{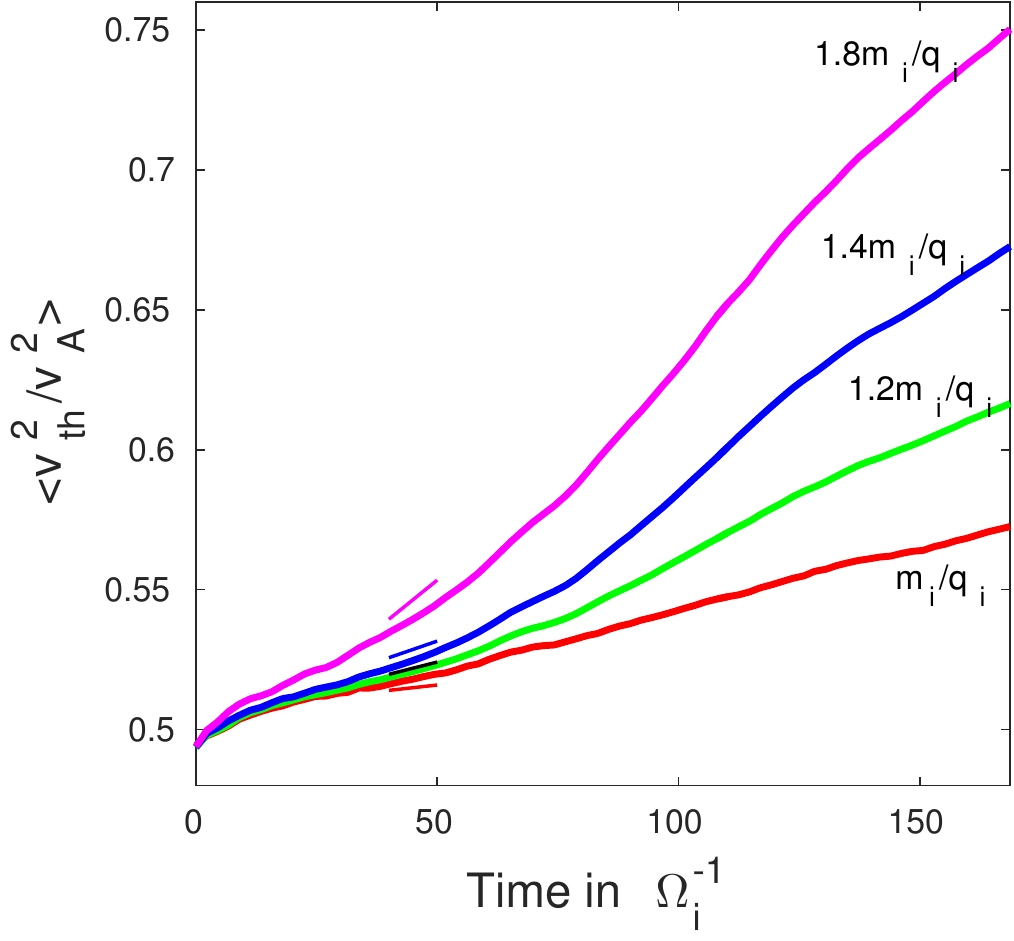}
  \caption{The red, green, blue and magenta curves show temporal evolution of mean square velocity of particles with mass to charge ratios 1, 1.2 ,1.4, and 1.8 times that of the simulation ions.  Alfven crossing time $\tau_A=128 \, \Omega_i^{-1}$. The small line segments of same color show analytical estimate (see section 4.1) of the instantaneous rate of heating for each species. Clearly, heavier particles are heated at a faster rate. }  
\label{v2}
\end{figure} 

\section{Heating and Acceleration}
The turbulent cascade generates electric and magnetic field perturbations from injection to kinetic scales of the plasma. Perturbations at certain dissipative scales are absorbed by the ions and electrons and converted into heat. The dissipation process may depend on several physical properties of the ion species, such as their charge, mass, and temperature. We have therefore placed several ion species in the simulation box to study dependence of heating rate on charge to mass ratios of the ions. Figure \ref{v2} shows the evolution of mean thermal speed of particles of differing charge to mass ratio as a function of time in units of ion cyclotron time $1/\Omega_i $. The thermal velocity of each particle is defined as $\vec{v}_{th}=\vec{v}- \vec{E} \times {B}_0 \hat{x}/ B_0^2$, where electric field $\vec{E}$ is the local value of the electric field interpolated from the grid and $\vec{B}_0=B_0 \hat{x}$. As evident from the figure, the heating rate strongly depends on the the Q/M of ion species. The thermal energy per nucleon for the ions of smaller Q/M increases at a much faster rate than that of the ions of larger Q/M. All the ion species shown in figure \ref{v2} were initially at the same thermal velocity. Although, the rate of heating marginally depends on the initial temperature of the ions, the trend of stronger heating rate for smaller Q/M particles persists. Additionally, equal thermal velocity criteria clearly identifies the differences in the heating rate solely due to differences in the cyclotron frequency of the ion species.

\begin{figure}[h]
  \includegraphics[width=0.48 \textwidth]{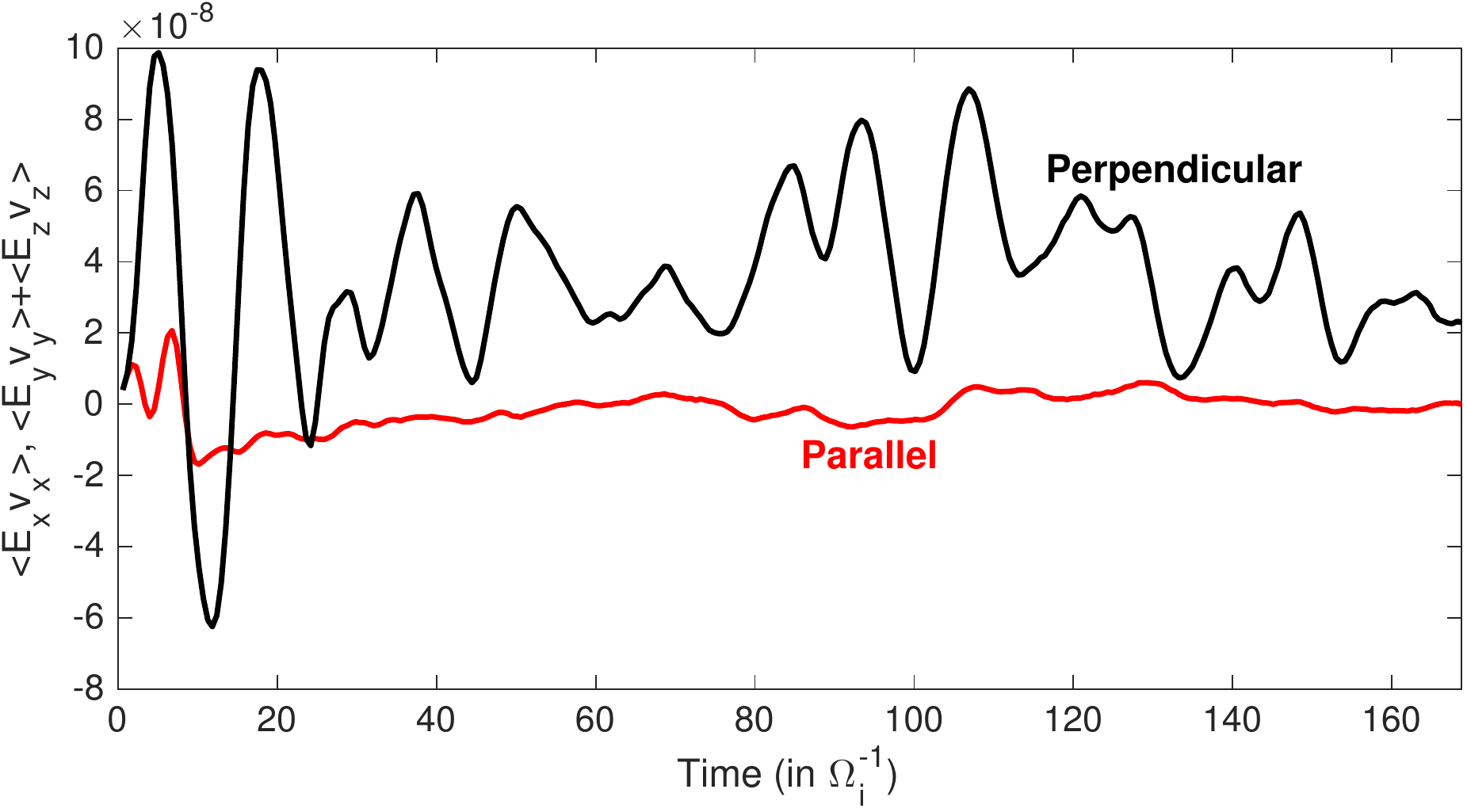}
  \caption{ The instantaneous net work done on all particles $(Q/M=0.5 q_i/m_i )$ by parallel and perpendicular components of the electric field is shown by red and black curves, respectively. Heavier ions are heated preferentially in the direction perpendicular to the background magnetic field.}  
\label{perp_heating}
\end{figure}

\begin{figure}[h]
  \centering
\subfloat[ ]{%
  \includegraphics[width=.48\linewidth]{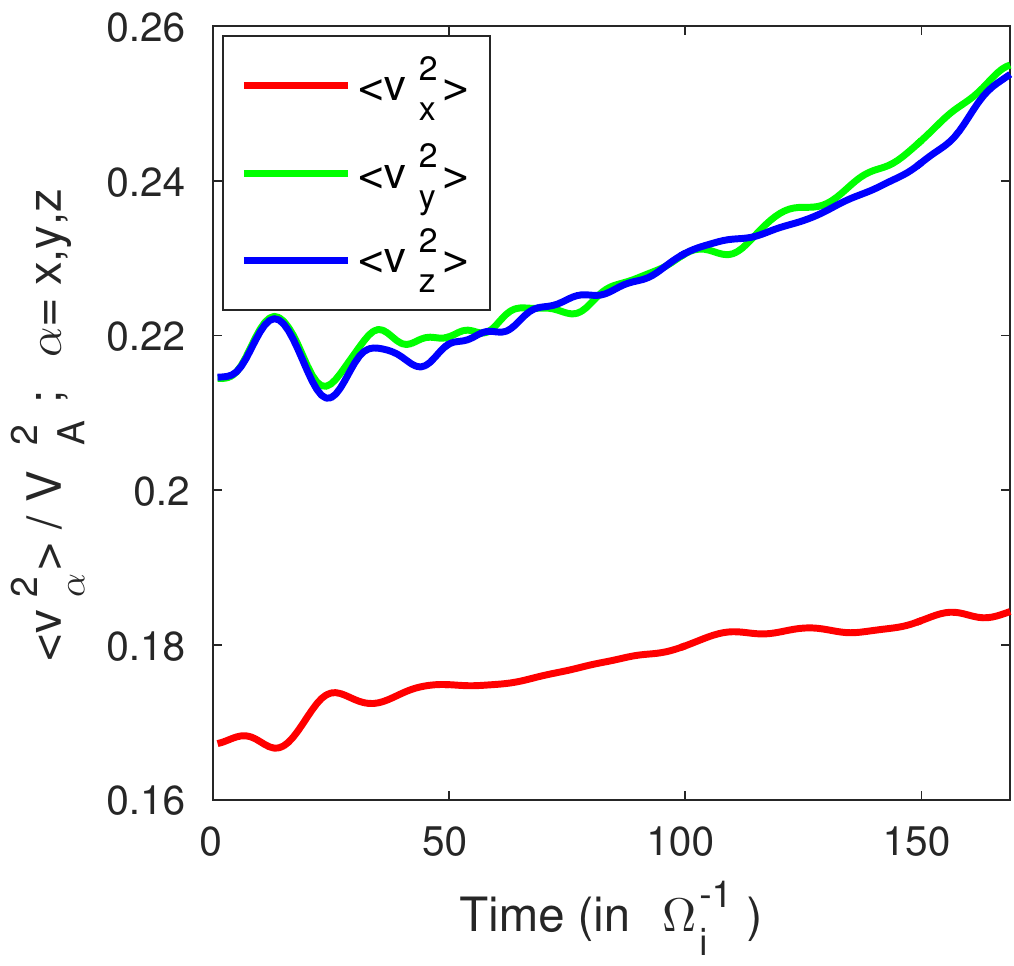}%
}\hfill
\subfloat[ ]{%
  \includegraphics[width=.49 \linewidth]{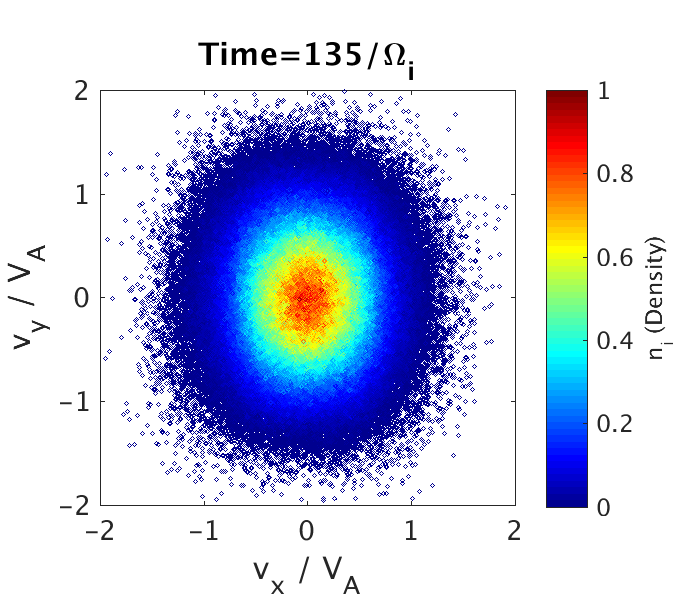}
 }\hfill
  \caption{ Left panel: Temporal evolution of the mean velocity square along three orthogonal axis are shown for heavy ion particles with $M/Q=1.8 m_i/q_i$.  Right Panel: An instantaneous location of ions of the same specie in two-dimensional velocity-space is shown as $v_x$(parallel) versus $v_y$(perpendicular). The color of each point in the map represents local density of the ions $N_i$ normalized by the maximum density. The distribution of heavy ions in phase-space develop small but appreciable anisotropy as a consequence of anisotropic heating (see figure \ref{perp_heating}). }  
\label{anisotropy}
\end{figure}

The heating of heavier ion species is anisotropic in momentum space (see also \cite{Chandran2010,Eliot2013,Vasquez2015}). In figure \ref{perp_heating} we show the instantaneous net work done on all particles by  parallel and perpendicular components of the electric field. It shows that most of the heating of heavy ions is due to perpendicular component of the electric field. Notably, the net work done by the parallel component of the electric field is marginally negative which could possibly be due to time dependent compressible modes with electric field along the magnetic field \citep[see also][]{Li1999,Hollweg1999}.  Anisotropic heating of the heavy ions leads to an anisotropic temperature of the ions where the perpendicular temperature of the ions is larger than the parallel temperature. The development of temperature anisotropy is illustrated in figure \ref{anisotropy} for a heavy ion specie. 

We suggest that cyclotron resonance with Alfven waves is the dominant mechanism for the heating of ions of smaller Q/M and hence of smaller gyro-frequency. The perpendicular heating of the ions and a strong dependence of heating rate on Q/M are both consistent with the resonant heating since the electric field of Alfven waves are in the transverse direction and power in the resonant wave is a steeply increasing function of the M/Q. In the following we estimate the rate of heating of heavy ions due to cyclotron resonance with Alfven waves.

\subsection{Estimate of Cyclotron resonant heating rate} 
The magnetic perturbations lead to random scattering of the pitch angle while the electric fluctuations stochastically alter the energy of the particles. Individual ions of charge to mass ratios Q/M gyrating in large scale background magnetic field $B_0$ can resonantly be accelerated or decelerated by the the similarly rotating transverse electric field $E_\perp$ of Alfven waves of frequency $\omega$ propagating along the magnetic field. The resonance condition is given by 
\begin{equation}
  \omega \pm k_\parallel v_{\parallel} = n \omega_c 
\label{resonance}
\end{equation}
where n is a positive integer and $\omega_c$ is the cyclotron frequency ($\omega_c=QB_0/Mc$). If the resonance takes place at a uniformly random angle between the electric field and the transverse component of the velocity of the particle, the particle would make random walk in the momentum space and the net energy of the ensemble of particles following Maxwellian distribution would diffusively increase with time. 

For the purpose of estimating the rate of heating we consider interaction of individual particles with turbulent electric and magnetic field which is considered to be an ensemble of shear Alfven waves, although the properties of the ensemble itself is governed by the collective motion of charged particles in the plasma. The wave particle interaction picture is more justified for the species occurring in trace amount or barely altering the turbulent dynamics, which is the case for the heavy ions in the simulations which are treated as the test particles. The change in the squared velocity of a particle (a measure of energy per unit mass) during a time interval  $\Delta t$ is given by

\begin{equation}
\Delta v^2  =   2 Q \vec{E}_\perp \cdot \vec{v}_\perp \Delta t /M  = 2 Q (\delta \vec{u}_\perp \times \vec{B}_0) \cdot \vec{v}_\perp/ M
\end{equation}
where $\delta \vec{u}_\perp$ is the local transverse drift velocity of all simulation particles. Considering a Fourier series representation of $\delta u_\perp$ we write  

\begin{equation}
\Delta v^2  = 2 Q/M \, (\sum_{\vec{k}} \delta \vec{u}_\perp(k_x,k_y,k_z) \times \vec{B}_0) \cdot \vec{v}_\perp
\end{equation}
where sum is over all discrete three dimensional wave vectors in the simulation box. In the case of a sharp resonance only the modes satisfying the resonant condition described by equation \ref{resonance} contribute to the a net change in the particle's thermal speed. Therefore, we restrict the sum to wave vectors satisfying the resonant condition, i.e. $k^{res}_\parallel ( V_A \pm v_{\parallel} ) =n\omega_c$.  We now define a dimensionless diffusion rate as, $D =  <(\Delta v^2_{th}/V^2_A)^2>/\Omega_i\Delta t  $, where the brackets refer to an average over all particles constituting an ion species. The average is estimated as in the Section 3 of \cite{David2014}, which gives 

\begin{equation}
   \begin{split}
      D  \simeq & \frac{2\pi Q (1+v_\parallel/V_A)}{ M}   \times \sum_{k_y, k_z}  \frac{J_1(M k_r v_\perp/ QV_A) }{k_r^2}  \\
       & \times (B^2_y(k^{res}_\parallel,k_y,k_z)+B^2_z(k^{res}_\parallel, k_y,k_z))/B_0^2
       \end{split}
\end{equation} 
where $k_r=c\sqrt{k_y^2+k_z^2}/\omega_{pe}$, and $J$ stands for Bessel functions of the first kind. 
Here we have considered only $n=1$ resonance and eddy turn over time at the length scale $1/k_\parallel$ is taken to be $k_\parallel V_A$.  The first term in the summation indicates that only a fraction of particles whose Larmor radius is smaller than the perpendicular wavelength of the oblique modes are efficiently energized due to that particular oblique mode. 

The above expression for the diffusion rate can be used to obtain an estimate of the heating rate due to cyclotron resonance for any species in the simulation. However, until one Alfven crossing time the net magnetic power in the resonance modes for the ions species shown in figure \ref{v2} is increasing with time as the turbulent energy cascades down from the larger length scales. Additionally, increasing parallel velocity of the ions due to heating also enhances the rate of heating at later times. Therefore, the diffusion rate is increasing with time and the rate of heating in the simulation must be obtained from a time dependent diffusion in momentum space. We use the mean value of  $k^{res}_\parallel$, $v_\parallel$, and $v_\perp$ to estimate the value $D$ for each species as a function of time. Taking $D=D_0 t $ within a time interval $\Delta t$ we get $<\Delta v_{th}^2/V^2_A> \propto \Delta t$. The instantaneous heating rate thus obtained at $ t=45/\Omega_i $ is shown in figure \ref{v2} as straight lines. The analytically estimated rate of heating due to cyclotron resonance appears to be consistent with the observed rate of heating of the heavy ions.

\begin{figure}[h]
  \centering
  \includegraphics[width=0.4\textwidth]{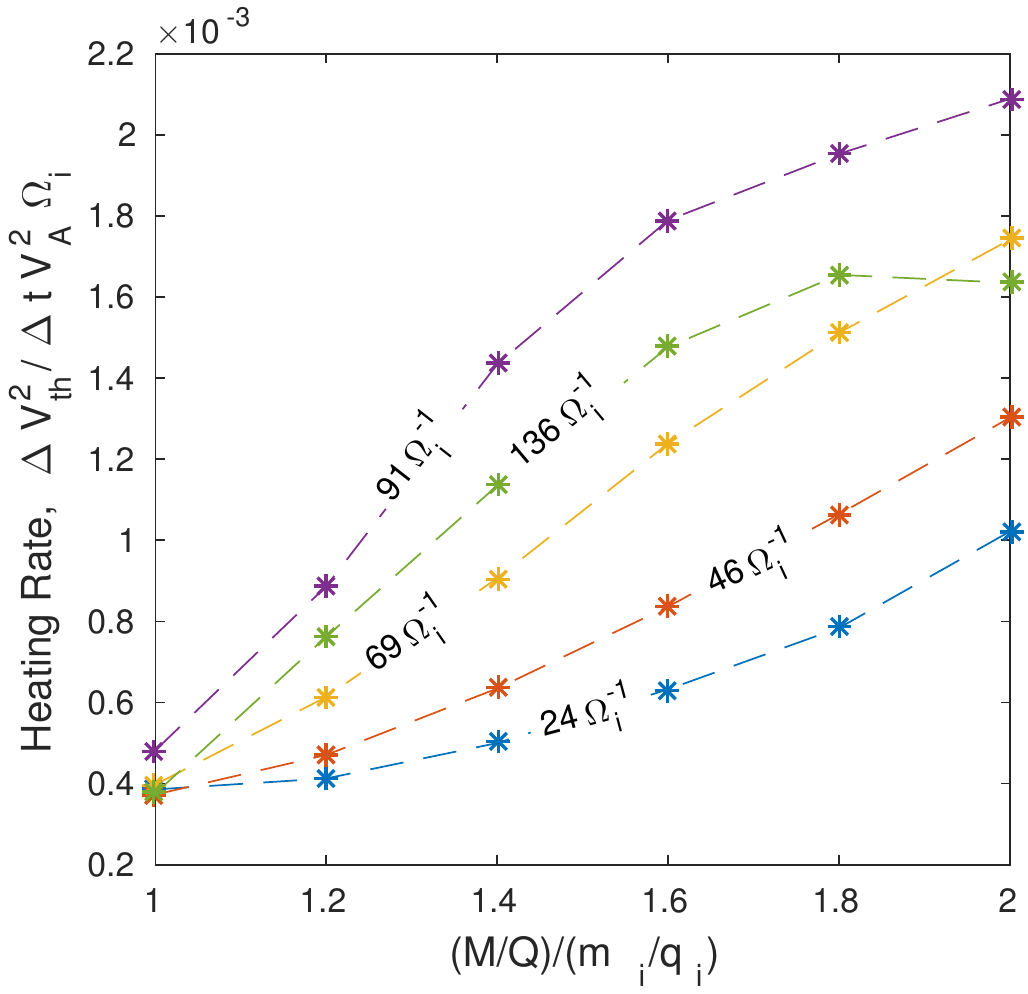}
  \caption{ The rate of change of mean velocity square is shown for species of decreasing charge to mass ratios. Each set of data points of same color connected by a dashed line correspond to different stage of the simulation labeled by the time from the beginning of the simulation. All species shown here had same initial thermal speed.}  
  \label{heating_rate}
\end{figure}

In figure \ref{heating_rate} we show the rate of heating at different stages of the simulation for heavier on species of decreasing charge to mass ratio as compared to the simulation ions. Up until about one Alfven crossing time the spectrum of magnetic fluctuation is steep and ions of larger M/Q are preferentially heated at a faster rate. The Alfven waves that ions can resonate with have a parallel wavelength of $\lambda_{res} \sim 2\pi (c/\omega_{pe}) \times (1+v_\parallel/V_A)\times M/Q$. At later stages the magnetic power in large scale waves declines and the spectrum tends to be flatter (see figure \ref{spectrum}). Consequently, the ions ($Q/M \lesssim 0.5 q_i/m_i $ ) resonating with the decaying flatter part of the spectrum are heated at slower rate and cease to enjoy a preferential heating. 

\subsection{Non Thermal Acceleration}

\begin{figure}[h]
  \includegraphics[width=0.48 \textwidth]{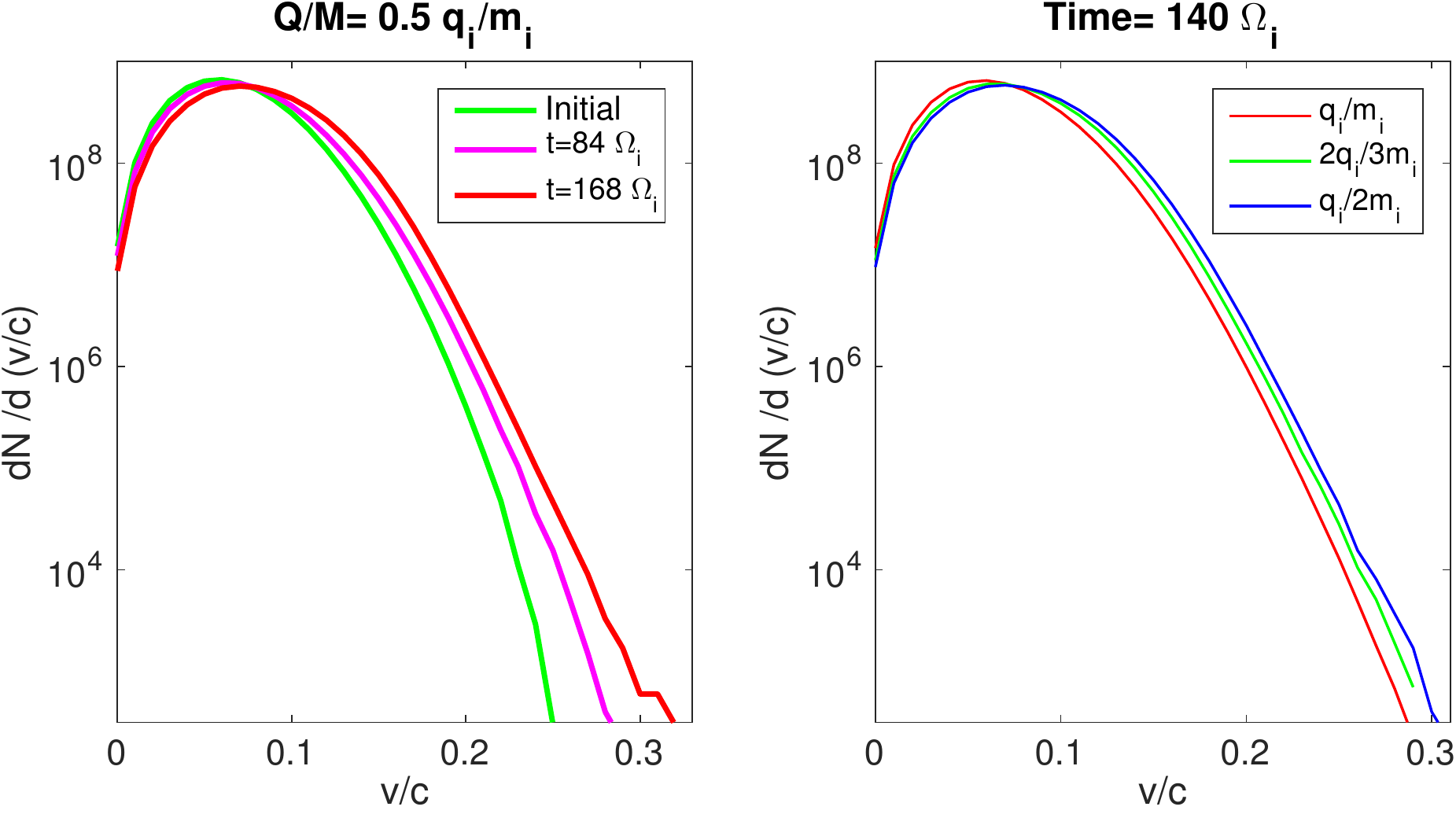}
  \caption{Left panel: The distribution of speed at t=0, 84, and 168 $\Omega_i^{-1}$ are shown by green, magenta, and red curves, respectively, for particles of charge to mass ratio $0.5 q_i/m_i$ . Right Panel: Velocity distribution at time $140 \Omega_i^{-1}$ is shown for species with $(Q/M)/(q_i/m_i)$=1, 2/3, and 0.5 by red, green, and blue curves, respectively. Initially the velocity distribution is a drifting Maxwellian which develops a non-thermal tail at later times. }  
\label{vel_spec}
\end{figure}

The resonance condition in equation \ref{resonance} implies that particles moving faster along the magnetic field lines are able to resonate with Alfven waves of smaller frequency or smaller wave number. Therefore, rate of gain in energy is an increasing function of particle's speed which can lead to acceleration of particles to very high energies. The time evolution of the distribution of speed is illustrated in figure \ref{vel_spec}. It is clear from the left panel that the velocity distribution, which is initially a drifting Maxwellian, develops a non-thermal tail at later times.  

In figure 6, we show the history of mean parallel and perpendicular velocity of few selected particles that become significantly more energetic than the particles constituting bulk of the thermal population. It is clear that these selected particles had significantly larger speed to begin with and were at the tail of initial Maxwellian distribution.   

\begin{figure}[h]
  \includegraphics[width=0.48 \textwidth]{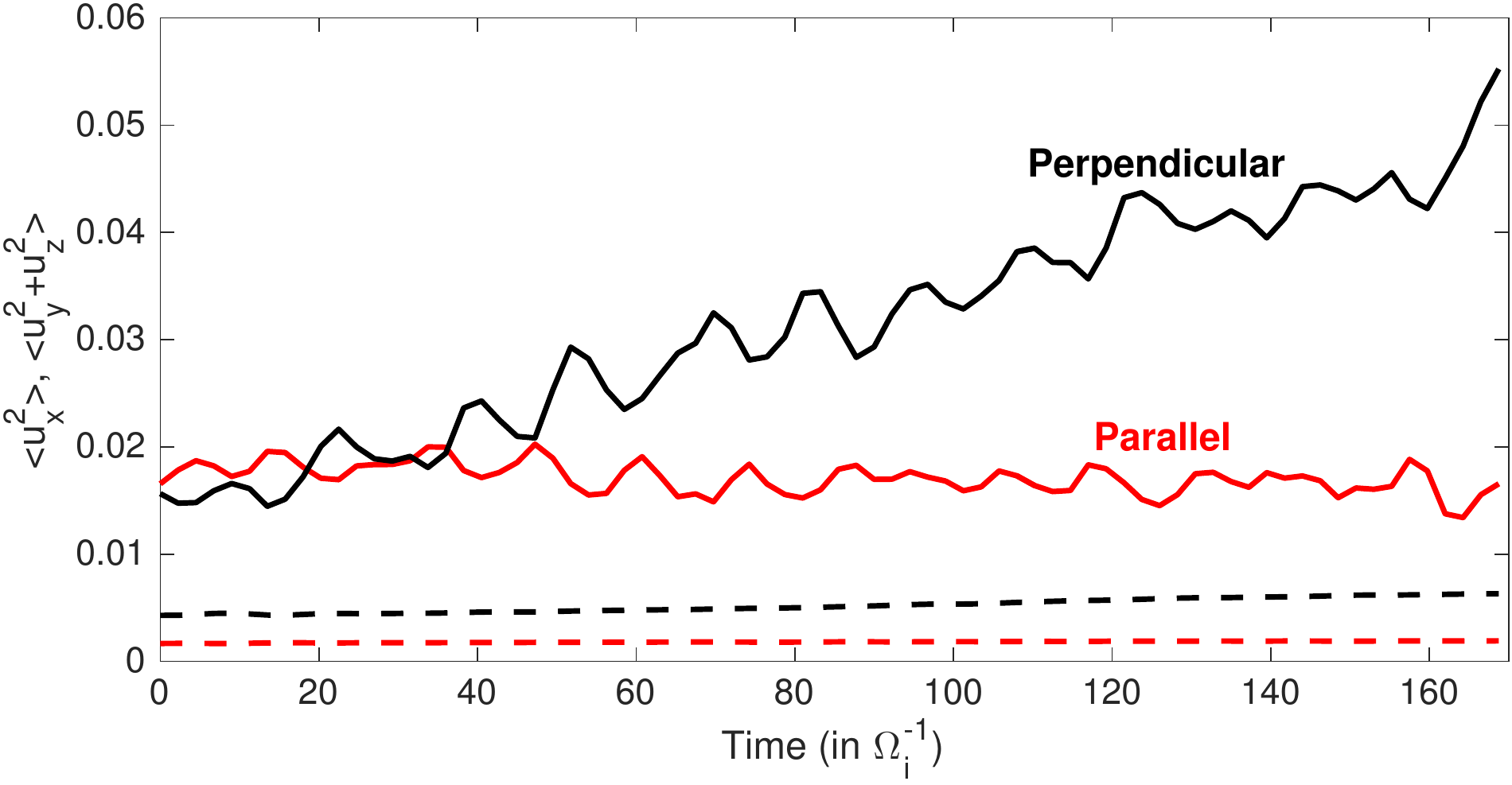}
  \caption{ The red and black curves show mean square of parallel (along x-axis ) and perpendicular components of particle's three-velocity, respectively, as a function of time for a species of $Q/M=0.5  q_i/m_i$. The solid lines show the average quantities (instantaneous particle average) for few selected particles that constitute non-thermal tail ($v> 0.25 c=2.5 V_A$) of the velocity distribution at t=$168\Omega_i^{-1}$ . For comparison, the dashed curves show the same averages over all particles. Particles that are accelerated to non-thermal energies are the one with larger parallel velocity.}
\end{figure}

\section{Discussion and Conclusions}
We have simulated turbulence cascade in a magnetized plasma using fully electromagnetic PIC method which is suitable for modeling kinetic physics of all plasma species. The relative rate of absorption of turbulent energy trickling down from larger scales by particles of different charge to mass ratios is studied by simultaneously evolving all particles in the phase-space. We find that the plasma species of smaller charge to mass ratios are heated and accelerated at a significantly higher rate. We have suggested that the heating and acceleration of these heavier ions are predominantly due to cyclotron resonance with Alfven waves. 

The physical size of the simulation box limits the extent of inertial scale in the simulation which determines the range of value of Q/M of heavy ions that are preferentially heated and accelerated. The range of mass to charge ratios and the temporal and spatial scales involved in impulsive SEPs is much larger than the scales simulated here. However, the rate of heating of heavy ions observed in the simulation is consistent with our analytical estimates of the heating expected due to the cyclotron resonance. In impulsive SEPs where the resonance can take place over rather extended inertial range of the turbulence the cyclotron resonance would imply large enhancement of even ultra-heavy ions.

The sources of Impulsive SEPs are not entirely known but their association with Type III ratio bursts \citep{Wang2012} has led to identification of their origin as extended regions in the solar corona associated with coronal jets \citep{Nariaki2015}. The variation in the energy spectrum of particles and enhancement ratios which does not show any apparent correlation with the properties of the sources suggest that the observed heating may take place over a wide range of plasma parameters, which favors a rather robust physical process viable over varying source conditions. The observed features of $^3He$ rich SEPs appear to be consistent with the generic nature of the heating and acceleration in anisotropic Aflvenic turbulence. However, the rate of heating via cyclotron resonance strongly depends on the spectrum of Alfvenic fluctuations and the extent of inertial range of the turbulence which determines total power in the Alfven waves at the resonate scale of the particles. A steeper spectrum enhances differentiation in the resonant heating of particles of varying M/Q. On the other hand, a steeper implies smaller power in the resonant modes at the kinetic scales and therefore a larger heating and acceleration time for all plasma species. For an estimated best fit temperature of $~10^7 K$ at the source of impulsive SEPs, preferential acceleration is observed for nuclei with $M/Q \gtrsim 10$. In case of acceleration due to the turbulence it would imply that the scale at which turbulence is isotopic is at least hundred times larger than the mean gyroradius of protons. The future observations, numerical simulations, and a better statistics of events might shed more light on the environments and mechanisms responsible for acceleration and heating of ions in impulsive SEPs.

The preferential heating of heavy ions is evident in other heliospheric environments as well, such as in Solar corona and fast solar wind\citep{Kohl2006,Cranmer2009,Gloeckler1995,Kasper2013,Tracy2015}. While some of enhancement of heavy nuclei could be due to other physical mechanisms, such as reacceleration in shocks, magnetic reconnections\citep{Drake2009}, and resonance with ion cyclotron waves \citep{Fisk1978, Drake2011}, cyclotron damping of Alfvenic turbulence may still be playing an important or possibly a dominant role \citep{Cranmer2014} in heating and acceleration process.

We thank M. Kunz, A. Philippov, and A. Schekochihin for helpful discussions. We acknowledge support from the Israel-U.S. Binational Science Foundation, the Israeli Science Foundation (ISF), the ISF-University Grant Commission (India), NSF grant AST-1517638, and the Joan and Robert Arnow Chair of Theoretical Astrophysics(D.E.). M.G. is supported by NASA through Einstein Postdoctoral Fellowship Award Number PF-160137 issued by the Chandra X-ray Observatory Center, which is operated by the Smithsonian Astrophysical Observatory for and on behalf of NASA under contract NAS8-03060. Simulations in this paper used computational resources supported by PICSciE-OIT High Performance Computing Center and the NASA/Ames HEC Program (SMD-16-6751; Pleiades).

\bibliographystyle{apj} 
\bibliography{alfven_turb}

\end{document}